\documentclass{aa}
\usepackage{graphicx} 
\usepackage{subcaption}

\usepackage{caption}
\usepackage{float}
\usepackage{txfonts}
\usepackage{siunitx}
\usepackage{xspace}
\usepackage{float}
\usepackage[switch]{lineno}
\usepackage[colorlinks,allcolors=blue]{hyperref}
\usepackage{multicol}
\usepackage{multirow}

\makeatletter
\renewcommand*\aa@pageof{, page \thepage{} of \pageref*{LastPage}}
\makeatother

\begin{document}

\title{A novel approach to optimizing the image cleaning performance of Imaging Atmospheric Cherenkov Telescopes: Application to a time-based cleaning for H.E.S.S.}

\titlerunning{Optimizing the image cleaning performance of Imaging Atmospheric Cherenkov Telescopes}

\author{
  Jelena \'{C}eli\'{c}\inst{\ref{ECAP}} \and 
  Rodrigo Guedes Lang\inst{\ref{ECAP}} \and
  Simon Steinmassl\inst{\ref{MPIK}} \and
  Jim Hinton\inst{\ref{MPIK}} \and 
  Stefan Funk\inst{\ref{ECAP}}
}
\authorrunning{J.\'{C}eli\'{c} at al.}
\institute{
Friedrich-Alexander-Universit\"at Erlangen-N\"urnberg, Erlangen Centre for Astroparticle Physics, Nikolaus-Fiebiger-Str. 2, 91058 Erlangen, Germany\\ \email{jelena.celic@fau.de} \label{ECAP} \and
Max-Planck-Institut f\"ur Kernphysik, Saupfercheckweg 1, 69117 Heidelberg, Germany\\ \label{MPIK} 
}

\abstract{The Imaging Atmospheric Cherenkov Telescope (IACT) technique is essential for gamma-ray astronomy, but it suffers from performance degradation due to night sky background (NSB) noise. This degradation is mitigated through image-cleaning procedures. This study introduces a time-based cleaning method for the High Energy Stereoscopic System using CT5 in monoscopic mode, and it presents an optimization workflow for image-cleaning algorithms to enhance telescope sensitivity while minimizing systematic biases. 
Unlike previous optimization efforts, we do not use first-order metrics such as image size retention; instead, our pipeline focuses on the final sensitivity improvement and its systematic susceptibility to NSB.
We evaluate three methods –tail-cut cleaning and two variants of time-based cleaning, \texttt{TIME3D} and \texttt{TIME4D} –and identify the best-cut configurations for two cases: optimal overall sensitivity and minimal energy threshold. The \texttt{TIME3D} method achieves a $\sim 15\%$ improvement compared to standard tailcut cleaning for $E < 300$ GeV, with a $\sim 200\%$ improvement for the first energy bin (36.5 GeV < E < 64.9 GeV), providing more stable performance across a wider energy range by preserving more signal. The \texttt{TIME4D} method achieves an approximately$\ 20\%$ improvement at low energies due to superior NSB noise suppression, allowing enhanced capability to detect sources at the lowest energies.
We demonstrate that using first-order estimations of cleaning performance, such as image size retaining or NSB pixel reduction, cannot provide a complete picture of the expected result in the final sensitivity. Beyond expanding the effective area at low energies, sensitivity improvement requires precise event reconstruction, including improved energy and directional accuracy. Enhanced gamma-hadron separation and optimized pre-selection cuts further boost sensitivity. The proposed pipeline fully explores this, providing a fair and robust comparison between different cleaning methods. The method is general and can be applied to other IACT systems, such as the Very Energetic Radiation Imaging Telescope Array System and the Major Atmospheric Gamma-Ray Imaging Cherenkov Telescopes. By advancing data-driven image cleaning, this study lays the groundwork for detecting faint astrophysical sources and deepening our understanding of high-energy cosmic phenomena.}

\date{\today}


\maketitle

\section{Introduction}\label{sec1}
The Imaging Atmospheric Cherenkov Telescope (IACT) technique, first demonstrated by Whipple in 1989, has proven extremely successful in detecting gamma rays in the energy range from 100 GeV to several tens of TeV~\citep{Weekes1989}. Over the past decades, modern IACT systems, such as the High Energy Stereoscopic System (H.E.S.S.) ~\citep{HessCrab2006}, Major Atmospheric Gamma-Ray Imaging Cherenkov Telescopes (MAGIC)~\citep{MAGIC:2011Crab}, and the Very Energetic Radiation Imaging Telescope Array System (VERITAS)~\citep{VERITAS:2006lyc} have revolutionized our understanding of energetic processes in the Universe~\citep{Funk_2015}. The next-generation instrument is the Cherenkov Telescope Array Observatory (CTAO), which features significantly enhanced hardware and software capabilities. With a sensitivity up to ten times greater than that of existing instruments, CTAO is expected to enable groundbreaking discoveries \citep{ScienceWithCTA2018}. 

The detection principle of IACTs is based on observing gamma rays indirectly through extensive air showers. Secondary charged particles traveling faster than the speed of light in the atmosphere emit light in the form of Cherenkov radiation, which is collected by the telescope mirrors, forming an image of the air shower~\citep{Weekes:2005da}. From these images, it is possible to reconstruct the essential properties of the primary particle, such as its energy, direction, and species. 

However, Cherenkov images are affected by noise induced by the night sky background (NSB). Even on the darkest nights, faint diffuse light from various sources (such as night airglow, zodiacal light, or diffuse light from unresolved stars) contributes to the NSB~\citep{1998A&AS..127....1L, Preuss:2001za}. These components vary with time, location, and observing conditions, making the NSB a dynamic and complex phenomenon. In addition, electronic and detector noise further contaminates the recorded images. This combination of noise sources introduces unwanted information into the event images, creating noise pixels that bias both image parameterization and the reconstruction of event properties.

To maximize the air shower signal while minimizing NSB contamination, robust image-cleaning techniques are essential for accurate data analysis. 
Traditional methods apply a pixel amplitude threshold to eliminate low-intensity pixels (\cite{Punch:1994sn},\cite{KONOPELKO1996199}), or use an island-cleaning technique to exclude isolated clusters~\citep{BOND2003311}, in order to retain Cherenkov light while reducing the number of noise pixels. With the development of advanced detector hardware, experiments such as MAGIC~\citep{ALIU2009293}, VERITAS~\citep{Maier:2017wzr}, and CTAO~\citep{CTAMaxim} have explored a different approach by incorporating timing information from the signal. The motivation for using timing information lies in the correlated temporal patterns of shower pixels, which contrast with the random temporal nature of NSB noise. By analyzing both spatial and temporal patterns, these techniques are expected to suppress NSB noise more effectively while preserving the Cherenkov signal. This approach is particularly advantageous for detecting low-energy, faint showers, ultimately lowering the system's energy threshold by retaining more low-energy events.

To address these challenges, we introduce a novel time-based cleaning technique developed for the H.E.S.S. system, as time information can now be exploited due to the last camera upgrade. Additionally, we propose a new workflow for optimizing image cleaning performance. The primary goal of this workflow is to improve the final detector sensitivity while making the system less susceptible to NSB fluctuations. Although the workflow is demonstrated using the H.E.S.S. system, it is designed to be applied to any IACT system, providing a robust and flexible solution for investigating current and future image-cleaning methods.

\section{Dataset description and image-cleaning algorithms}
The High Energy Stereoscopic System (H.E.S.S.) is an array of IACTs located in Namibia. The array initially comprised four telescopes, each with a 12-meter-diameter mirror (CT1 to CT4), which have been operating since 2003. In 2012,  a larger telescope was added with a 28-meter-diameter mirror in the center (CT5), which operates with a monoscopic trigger mode \citep{vanEldik2016}. In this work, we use Monte Carlo (MC) simulations of gamma-ray events recorded with CT5, equipped with the FlashCam (FC) camera ~\citep{Bi:2021V5}. Air shower simulations were generated with the \texttt{CORSIKA} (COsmic Ray SImulations for KAscade)~\citep{Heck:1998vt} software package. Simulations of the telescope response were performed using the \texttt{simtel\_array}~\citep{Bernl_hr_2008} software, using the most recent configuration files~\citep{Leuschner:2023ega}. The simulated MC gamma-ray spectrum follows a power law with an index of -2 to maintain high statistical rates at higher energy. 
Using simulations, we can obtain additional information about the true Cherenkov signal\textbf{–}specifically, the pixels associated with the event signal\textbf{–} which will be used later in the analysis. To provide a realistic background, we use real observational data with masks for known gamma-ray sources and bright stars, rather than MC proton simulations. This approach minimizes discrepancies between Monte Carlo and real data (\cite{PhysRevD.100.023010}, \cite{Pastor_Guti_rrez_2021}), while also limiting the significant computational cost associated with generating sufficiently large proton datasets for high-statistics analyses. Simulations were performed under representative conditions of a 20 degree zenith, a 0 degree azimuth, and a 0.5 degree offset. Background observations with zeniths around 20 degrees were chosen.
Calibration and data processing were performed using the H.E.S.S. analysis program (\texttt{HAP}), one of the current state-of-the-art tools for processing data from the H.E.S.S. telescopes. The FC pulse reconstruction, fully described in \cite{pühlhofer2021scienceverificationnewflashcambased}, provides the signal amplitude in photoelectrons (p.e.) and the peak times in nanoseconds for each pixel. The latter is considered to be the arrival time of the Cherenkov photons.

\begin{figure*}[ht!]
\centering
\includegraphics[width=\textwidth]{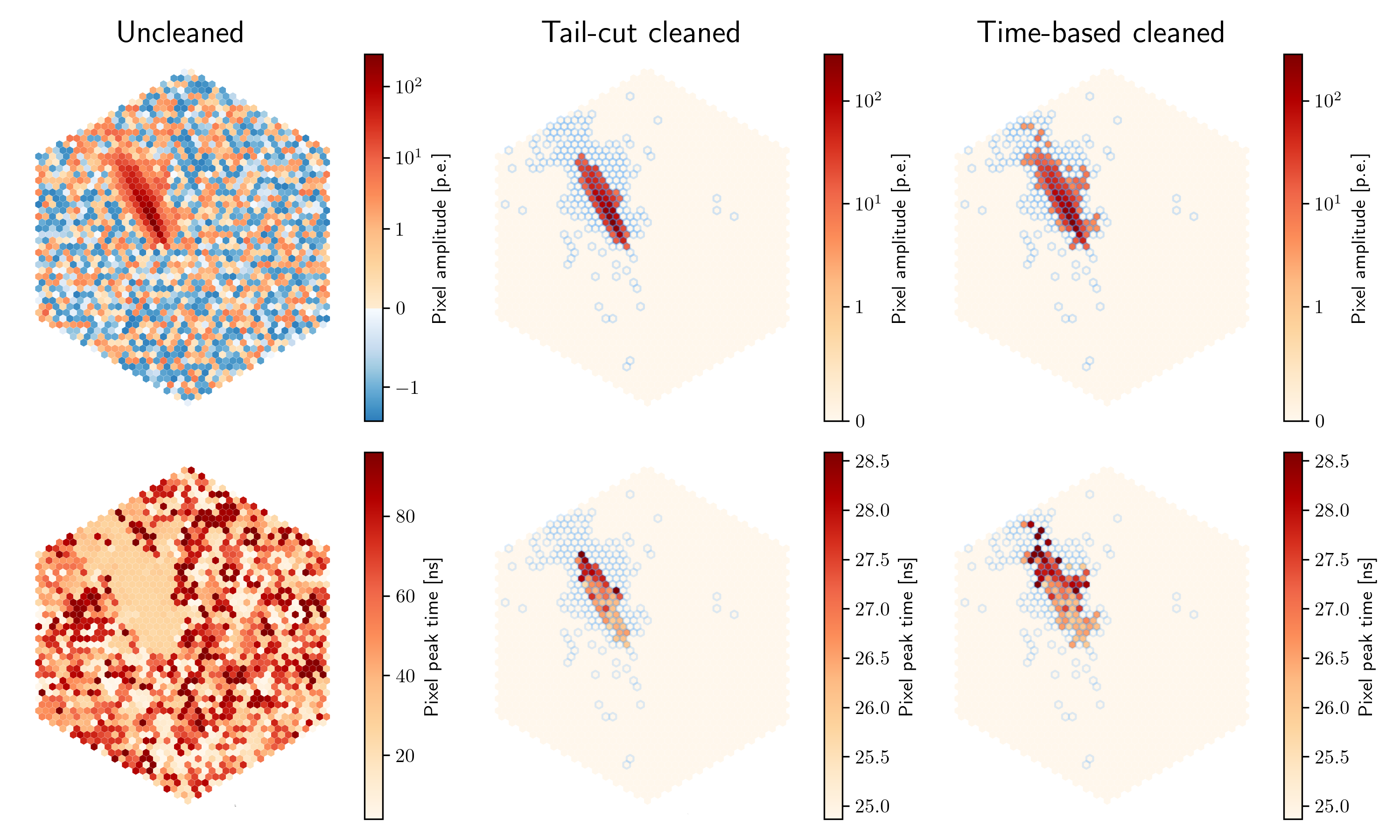}
\caption{An example MC gamma event seen before and after image cleaning with tail-cut and time-based algorithms. The amplitude (upper panel) and the peak time (lower panel) in each pixel of the FC are shown as  color scales. In the cleaned images, blue pixels indicate shower pixels associated with the Cherenkov signal, i.e., at least 1 p.e. comes from Cherenkov photons.}
\label{fig:CameraImages}
\end{figure*}

Once the signal in each of the pixels is reconstructed, image cleaning is performed.
The default image-cleaning method in HAP is tail-cut cleaning. In this work, we introduce a new approach, time-based cleaning for H.E.S.S.\textbf{–} designed to improve the separation of signal from noise by leveraging the timing characteristics of Cherenkov light. Cleaning differences are most impactful for events with low signal intensities, which arise from low-energy showers or large impact distances. While this affects all telescope types to some extent, the lowest energies\textbf{–}where intensities are most frequently near the cleaning threshold\textbf{–} are primarily accessible with CT5. In a stereo analysis, it is likely that one telescope will have a large impact distance, and thus a low signal. Nevertheless, the telescopes with the largest signal tend to dominate the reconstruction, diminishing the improved cleaning effect in the others. For these reasons, in this work we focus on a monoscopic analysis using CT5 FC, where the benefits of improved cleaning are more pronounced and directly traceable in the reconstruction performance.
The following sections provide a detailed description of the two cleaning algorithms. Figure~\ref{fig:CameraImages} illustrates the behavior of each algorithm, showing the remaining light\textbf{–} referred to in this work as size\textbf{–} after cleaning, for an example event.

\subsection{Tail-cut cleaning}
The standard image-cleaning technique used by H.E.S.S. is the two-threshold cleaning method, commonly referred to as tail-cut cleaning~\citep{HessCrab2006}. This method is favored for its effectiveness, simplicity, and computational efficiency. The cleaning process is executed in two stages, each applying specific thresholds to the pixel data. 
The noise level in each pixel is approximated by a Gaussian distribution with a mean of $0$,p and a standard deviation of $\sigma_{\rm{noise}}$. For observations under average Galactic NSB conditions with CT5 FC, a typical value is $\sigma_{\rm{galactic\,NSB}}$ = 1.65 p.e. \citep{FabiLeuschnerPhD}. In the current configuration, pixels with a signal below $3\sigma_{\rm{noise}}$ are removed in a pre-cleaning step. In the second stage, pixels with a signal above a specified threshold (e.g., 5 p.e.) are retained only if they are adjacent to a pixel exceeding a higher threshold (e.g., 10 p.e.). This condition is bidirectional: pixels above the higher threshold are retained if they are adjacent to a pixel with a signal above the lower threshold. This ensures that only regions with a significant signal are preserved, while isolated pixels, which are likely to be spurious signals or noise, are removed. Some cleaning configurations require multiple neighboring pixels to exceed the threshold, rather than just a single one. The default tail-cut cleaning configurations for CT5 are typically \texttt{0916NN2} and \texttt{0714NN2}. In these configurations, the first threshold is 9 p.e. or 7 p.e., the second is 16 p.e. or 14 p.e., and a minimum of two neighboring pixels (NN2) is required for a pixel to be retained. Thus, the usual tail-cut cleaning can be defined by four input parameters: $n_{\rm{noise}}$, $\rm{thr_1}$, $\rm{thr_2}$, and NN.

\subsection{Time-based cleaning using \texttt{DBSCAN}}

Traditional image-cleaning methods in IACTs, such as tail-cut cleaning, rely solely on pixel amplitudes to suppress background noise. Although effective in removing isolated noise pixels, these methods often fail to preserve the full extent of air shower images, particularly at lower energies. Low-energy and/or large-impact showers are characterized by faint Cherenkov light emissions, making them highly susceptible to removal by aggressive threshold cuts \citep{BOND2003311}. This leads to a significant loss of potentially valuable gamma-ray events, ultimately raising the energy threshold of the telescope. To overcome these limitations, time-based cleaning introduces an additional layer of information: the temporal evolution of the Cherenkov signal. The fundamental advantage of this approach is that shower pixels exhibit strong time correlations, whereas noise from the NSB follows a more random temporal distribution. By incorporating the arrival time of Cherenkov photons, time-based cleaning can distinguish real shower pixels from NSB-induced noise. By relaxing the amplitude cuts, more of the shower structure can be preserved while still effectively suppressing background contamination.

The time-based cleaning algorithm implemented in the HAP software uses density-based spatial clustering of applications with noise ( \texttt{DBSCAN}; \citep{10.5555/3001460.3001507}) to enhance the identification of clusters in IACT shower images. A detailed description of the implementation and first-order optimization of the input cleaning parameters can be found in~\cite{Steinmassl:2023fvf}. The optimized parameters derived in \cite{Steinmassl:2023fvf} are the current standard for CT5 FC and are labeled here as \texttt{TIME3D\_STD1} and \texttt{TIME3D\_STD2}; these are listed in Table \ref{tab:simons_parameters}.
\begin{table}[]
    \caption{Standard time-based cleaning parameter combinations of CT5 FC}
    \label{tab:simons_parameters}
    \centering
    \begin{tabular}{c|c|c}
         & \texttt{TIME3D\_STD1}  &  \texttt{TIME3D\_STD2} \\ \hline \hline
         $n_{\rm{hard}} \,$[p.e.] & 0 & 0\\
         $n_{\rm{noise}}\,[\sigma_{\rm{noise}}]$ & 3.0 & 3.5 \\
         minPts & 5 & 3 \\
         $s_{\rm{scale}} \,$[m] & 0.17 & 0.12\\
         $t_{\rm{scale}} \,$[ns] & 3.5 & 3.0\\ \hline\hline
    \end{tabular}
    \tablebib{(1)~\citet{Steinmassl:2023fvf}}
\end{table}
The \texttt{DBSCAN} algorithm is an efficient clustering method that operates without predefined cluster counts, making it particularly suited for IACT images. Hadronic showers often produce multiple clusters due to their interaction patterns, whereas gamma-ray showers typically result in a single, concentrated cluster. The algorithm's ability to classify unclustered points as noise further aids in isolating relevant data from background pixels. Before running \texttt{DBSCAN}, a pre-cut on intensity is applied, which can be either a fixed intensity threshold ($n_{\rm{hard}}$) or based on the noise level ($n_{\rm{noise}} \times \sigma_{\rm{noise}}$), similar to the pre-cleaning step of tail-cut cleaning. After pre-cleaning, \texttt{DBSCAN} is used to classify the remaining pixels. In general, the algorithm defines clusters based on two parameters: neighborhood distance $\epsilon$ and minimum points (minPts). 
Using these parameters, it classifies the points as core points (those with at least minPts within $\epsilon$), density-reachable points (within $\epsilon$ of a core point), or noise points (neither core nor density-reachable). Core points and their density-reachable neighbors are grouped to form clusters, while noise points are removed. 

The distance $\epsilon$ is defined as a dimensionless, scaled distance in an N-dimensional space. Two definitions are used: the first considers the pixel peak time and spatial position in the camera, resulting in a three-dimensional distance, and is labeled as \texttt{TIME3D}. The second adds the amplitude in the pixel as the fourth dimension by considering the logarithm of the ratio between the pixel amplitude and that of the brightest pixel; this is referred to as \texttt{TIME4D}. The distance between two pixels, $\epsilon_{ij}$, can then be expressed as

\begin{equation}
\begin{cases}
    \epsilon^{\rm{TIME3D}}_{ij} = \sqrt{ \left( \frac{x_i - x_j}{s_{\rm{scale}}} \right)^2 + \left( \frac{y_i - y_j}{s_{\rm{scale}}} \right)^2 + \left( \frac{t_i - t_j}{t_{\rm{scale}}} \right)^2 } \\ 
    \epsilon^{\rm{TIME4D}}_{ij} = \sqrt{ \left( \frac{x_i - x_j}{s_{\rm{scale}}} \right)^2 + \left( \frac{y_i - y_j}{s_{\rm{scale}}} \right)^2 + \left( \frac{t_i - t_j}{t_{\rm{scale}}} \right)^2 + \left( \frac{\log\left(A_i/A_0\right) - \log\left(A_i/A_0\right)}{A_{\rm{scale}}} \right)^2 },
\end{cases}
\end{equation}

where $x_i$ and $y_i$ are the spatial positions in the camera for pixel $i$, $t_i$ is its peak time, $A_i$ is its amplitude, and $A_0$ is the amplitude of the brightest pixel. The spatial, time, amplitude scales, $s_{\rm{scale}}$, $t_{\rm{scale}}$, and $A_{\rm{scale}}$ are input parameters of the cleaning (similar to the p.e. thresholds for tail-cuts). Thus, five input parameters ($n_{\rm{hard}}$, $n_{\rm{noise}}$, minPts, $s_{\rm{scale}}$, $t_{\rm{scale}}$)  must be optimized for \texttt{TIME3D} and six (additionally $A_{\rm{scale}}$) for \texttt{TIME4D}. Other possibilities for including amplitude information, such as those proposed by\cite{ESCANUELANIEVES2025103078}, were not investigated in this work.
To optimize computation, \texttt{DBSCAN} was implemented within the HAP framework using a k-d tree to precompute distances, thus reducing processing time.

\section{Optimization workflow}

\begin{figure*}[h!]
    \begin{minipage}[t]{.48\textwidth}
        \centering
        \includegraphics[width=\textwidth]{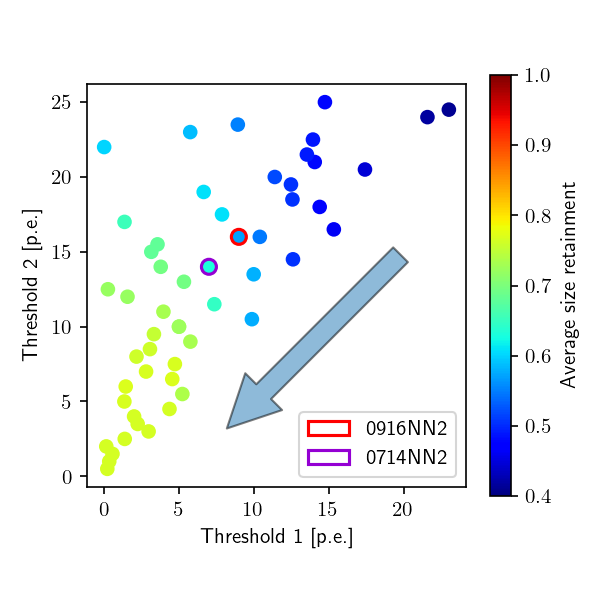}
        \subcaption{Average size retention}\label{fig:sizeretainment}
    \end{minipage}
    \hfill
    \begin{minipage}[t]{.48\textwidth}
        \centering
        \includegraphics[width=\textwidth]{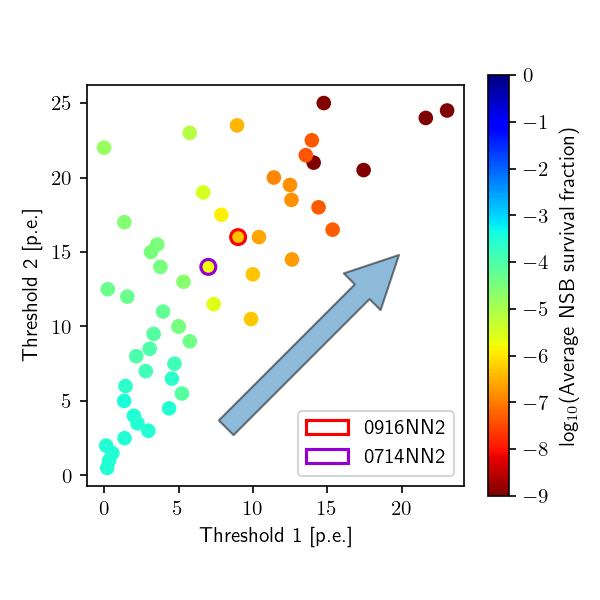}
    
        \subcaption{Average NSB survival rate}\label{fig:nsbsurvivalrate}
    \end{minipage}  
    \label{fig:1-2}
    \caption{Left: Average size retention, determined from pixels containing the shower signals. Right: Average NSB survival rate calculated from pixels not containing the shower signal. Both panels show different tail-cut threshold combinations. Red and purple circles  indicate the two default configurations. All combinations shown use the same pre-cleaning ($3\sigma_{\rm{noise}}$) as in the default configurations. Arrows indicate the direction of expected performance gain.}
\end{figure*}

The primary goal of image cleaning is to preserve as much shower-related information as possible while simultaneously reducing noise levels. However, a trade-off must be found. Figure \ref{fig:sizeretainment} shows the effect of varying thresholds in the tail-cut cleaning algorithm on the average remaining light (size) retention, calculated as follows:
\begin{equation}
R_{\text{size}} = \left\langle \frac{\sum\limits_{i \in P_{\text{ret}}} A_i}{\sum\limits_{i \in P_{\text{shower}}} A_i} \right\rangle.
\end{equation}

\noindent
Here, $R_{\text{size}}$ denotes the average size retention, i.e., the fraction of shower-associated light that survives cleaning.  $P_{\text{shower}}$ is the set of all pixels associated with the shower, as determined from Monte Carlo TruePE information, while  $P_{\text{ret}} \subset P_{\text{shower}}$ is the subset of those pixels that are retained after image cleaning. The variable $A_i$ denotes the reconstructed amplitude (in photoelectrons) of pixel $i$, including contributions from NSB and detector effects. The angle brackets $\langle \cdot \rangle$ indicate an average over all MC gamma-ray events.

The quantity is evaluated on shower-only pixels, which can be extracted from the simulations. The results indicate that as the thresholds increase, the size retention decreases. This means that higher thresholds result in more of the shower’s information being removed. Conversely, as thresholds are lowered, the number of NSB clusters increases, as shown in Figure \ref{fig:nsbsurvivalrate} where the average NSB survival rate rises significantly for lower thresholds. Finding the right balance between size retention and NSB cleaning efficiency is crucial to developing an optimal cleaning algorithm. For this reason, the optimization of image-cleaning methods remains a complex challenge. Previous studies \cite{Steinmassl:2023fvf}, \cite{CTAMaxim} and \cite{Kherlakian:2023xmz} aimed at improving image cleaning for IACTs have often focused on maximizing size retention, minimizing the number of survived NSB clusters, or combination of both. Although these are useful considerations, estimating the impact of image cleaning on the final sensitivity of the experiment is computationally expensive. Sensitivity depends on a complex interplay between preserving the gamma-ray signal, minimizing noise contamination, and maintaining accurate event reconstruction. Moreover, size thresholds for preselection are typically optimized for specific cleaning algorithms, which may not translate well across different methods, complicating direct comparisons.
Full reconstruction, separation training, and the instrument response function (IRF) generation must be performed for each cleaning using high statistics (see~\citep{unbehaun2025improvementsmonoscopicanalysisimaging} for a detailed description of the process). For that reason, it is impossible to repeat the process for hundreds or thousands of possible input cleaning parameters in order to find a global optimum. In response to this challenge, we established a new pipeline to efficiently explore cleaning configurations with respect to their impact on telescope sensitivity rather than on separate metrics. This approach provides a fair and robust method for comparing different cleaning algorithms.

The optimization framework developed in this study provides a generalized and data-driven approach for improving image cleaning in IACTs and consists of the following steps.  Candidate parameter sets are selected from the image-cleaning parameter space. The number of candidates is then reduced through a multistep cleaning process. First, an NSB susceptibility limit is applied, calculated from nominal and high NSB gamma simulations according to the formula \ref{eq:NSBsusceptibility formula}. For each candidate, the event size distribution of MC gamma and background events is generated and the shapes of both distributions are characterized using the fit function in \ref{eq:Fitfunction}. Candidates with similar behavior are then clustered using the K-means algorithm, and cluster representatives are determined as those closest to the cluster centers.

For all cluster representatives, the final sensitivity is calculated and the sensitivity improvement, $\xi$ (see \ref{eq:sens_improvement}), is evaluated for both performance and detection criteria. Candidates susceptible to NSB fluctuations, exceeding a fake cluster rate of 1$\%$, or with an effective area ratio test $\kappa$ above 1.1, are discarded.
 The optimal candidate is identified as the one with the highest sensitivity improvement (high $\xi$) and the greatest robustness to NSB fluctuations ($\kappa$=1). The first step in the proposed pipeline is to explore a wide range of input parameters for each of the three cleaning algorithms tested. 

\begin{table}
\caption{Chosen parameter spaces for testing the tail-cut cleaning algorithm.}
    \label{tab:tailcutpars}
    \centering
    \begin{tabular}{c|c}
         &  Chosen parameter spaces \\ \hline \hline
     $n_{\rm{noise}} \, [\sigma_{\rm{noise}}]$& 0, 1, 3, 5\\
     $\rm{thr_1}$, $\rm{thr_2}$ [p.e.] & 0.0 -- 25.0 \\
     NN & 1, 2, 3, 4, 5\\ \hline \hline
    \end{tabular}
    \tablefoot{The range notation used for parameters such as $\rm{thr_1}$ and $\rm{thr_2}$ indicates the upper and lower limits of a uniform distribution from which values are randomly drawn.}
\end{table}

\begin{table}
\caption{Chosen parameter spaces for \texttt{TIME3D} and \texttt{TIME4D}.}
    \label{tab:timecleaningpars}
    \centering
    \begin{tabular}{c|c}
         &  Chosen parameter spaces \\ \hline \hline
     $n_{\rm{hard}} \, [\rm{p.e.}]$    &  0, 3, 5 \\
     $n_{\rm{noise}} [\sigma_{\rm{noise}}]$    &  0.0 -- 7.2 \\
     minPts & 3, 5, 7, 9\\
     $s_{\rm{scale}}$[m] & 0.0 -- 0.4\\
     $t_{\rm{scale}}$[ns] & 0.0 -- 7.0 \\
     $A_{\rm{scale}}$ & -8 -- 0\\ \hline \hline
    \end{tabular}
    \tablefoot{The range notation used for parameters such as $n_{\rm{noise}}$, $s_{\rm{scale}}$, $t_{\rm{scale}}$, and $A_{\rm{scale}}$ indicates the upper and lower limits of a uniform distribution from which values are randomly drawn.}
\end{table}

Rather than performing a regular grid search, parameter values were randomly sampled to enable a more diverse and comprehensive exploration of the parameter space while significantly reducing the computational cost to a manageable level. The chosen parameter space limits are listed in Table \ref{tab:tailcutpars} and Table \ref{tab:timecleaningpars}. Establishing a broad coverage of the multidimensional parameter space increases the likelihood of identifying optimal configurations under various conditions.  
A parameter-space reduction strategy, described in the following section, was applied to remove less promising cleaning parameter configurations (hereafter referred to as candidates) and to focus the workflow on a smaller set of high-potential candidates. This approach enables the pipeline to prioritize sensitivity over individual metrics, providing a comprehensive and efficient manner of identifying the best cleaning methods and respective input parameters.

\subsection{Parameter-space reduction}

\subsubsection{NSB susceptibility as a pre-selection condition}
The first step in restricting the parameter space is to evaluate the susceptibility of each candidate to different NSB levels.  
This aspect is crucial because the response of the experiment must remain stable under varying NSB conditions, which can differ significantly depending on observation time, conditions, and pointings. A boosted decision tree (BDT) model with a size-dependent cut, as presented in \cite{unbehaun2025improvementsmonoscopicanalysisimaging}, was used. The BDT was trained to distinguish between MC gamma simulations under nominal NSB conditions and background data. For each candidate, the trained model was applied independently to MC simulations with both the nominal and elevated NSB levels. The NSB susceptibility was then estimated by the difference in gamma efficiency between the two NSB levels,

\begin{equation} \label{eq:NSBsusceptibility formula}
     \Delta \gamma_{\rm{eff}} = \sqrt{(\gamma^{\rm{nomNSB}}_{\rm{eff}}-\gamma^{\rm{highNSB}}_{\rm{eff}})^2}.
\end{equation}

The calculation was repeated one hundred times for each candidate, and the average susceptibility was determined. This repetition was necessary because the analysis relied on low-statistics datasets ($\approx 50000$ events) to reduce CPU time costs. 

\begin{figure}[h]
    \centering
    \includegraphics[width=\linewidth]{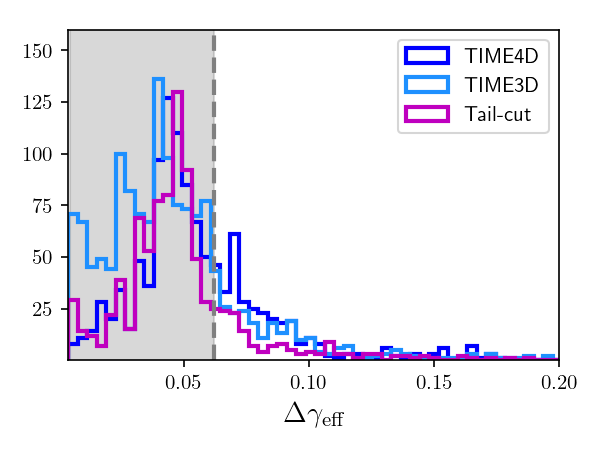}
    \caption{Distribution of differences in gamma efficiency for two NSB levels for the three tested cleaning algorithms: \texttt{TIME3D} (light blue), \texttt{TIME4D} (blue), and tail-cut (purple). The dashed line indicates the chosen upper limit ($6.2\%$). The discarded candidates, outside of the grey area, correspond to $13.4\%$ of all tested candidates.}
    \label{fig:nsbsus}
\end{figure}

Figure~\ref{fig:nsbsus} shows the distribution of NSB susceptibility. A value for $\Delta \gamma_{\rm{eff}}$ of $6.2\%$ was set as an upper limit, which is $10\%$ higher than the value for the standard \texttt{0714NN2} tail-cut cleaning. Applying this threshold resulted in $13.4\%$ of all tested candidate configurations being discarded. The total number of candidates per cleaning algorithm that survived this first step is listed in Table \ref{tab:nsb sus}.

\begin{table}
    \caption{Number of candidates before and after application of the NSB susceptibility cut.}
    \label{tab:nsb sus}
    \centering
    \begin{tabular}{c|c|c}
       Algorithm  & before NSB sus. & after NSB sus. \\
       \hline
       \texttt{TIME3D} & 2601 & 1147 \\
       \texttt{TIME4D}  & 2851 & 773\\
       Tail-cut & 1018 & 723
    \end{tabular}
\end{table}

\subsubsection{Clustering of the size distribution shape parameters}
After filtering the NSB-susceptible candidates, the second step in the optimization pipeline is to determine the size distributions of the cleaned events. Each cleaning method yields a unique size distribution that depends on its image cleaning settings and affects gamma rays and background data events differently. The distributions of both signal and background events are important. The effects on the former are reflected in the effective area (the energy-dependent area over which gamma rays are detected after all analysis cuts, reflecting both detection efficiency and reconstruction performance), as well as in the energy and angular resolution (defined as the 68\% containment radius of the reconstructed gamma-ray direction around the true source position, indicating directional precision) after reconstruction. The effect on the latter, which is often neglected in previous studies, directly impacts the gamma-hadron separation power. Specifically, a cleaning method that removes the differences between the gamma and hadron images will perform very poorly in terms of sensitivity. 

\begin{figure*}[h!]
    \centering
    \includegraphics[width=\linewidth]{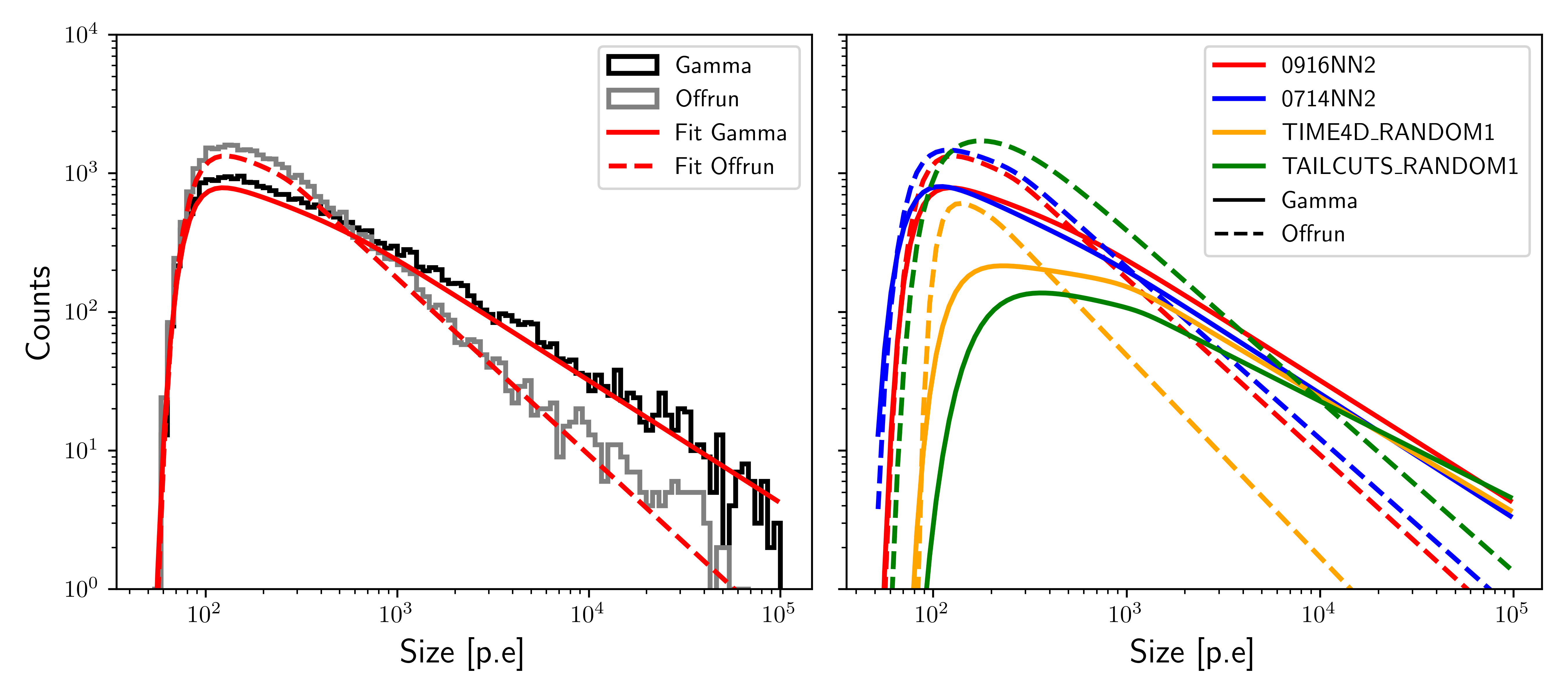}
    \caption{Left: Size distribution for gamma MC and off-data events, with corresponding fitted function for the default tail-cut candidate 0916NN2. Right: Fitted function for gamma MC (solid line) and off-data (dashed line) for different example cleanings (shown in different colors).}
    \label{fig:size_dist_cluster}
\end{figure*}

We propose an empirical fit function to describe the overall shape of the size distributions for gamma rays and background separately, using seven free parameters:

\begin{equation} \label{eq:Fitfunction}
    f(x) = A \cdot \left( \frac{x}{100} \right)^{g_1} \left( 1 + \left( \frac{x}{s_b} \right)^{\frac{1}{d_1}} \right)^{(g_2 - g_1) d_1}\cdot \exp\left( -\left( \frac{s_{\text{cut}}}{x} \right)^{d_2} \right),
\end{equation}

where $A$, $g_1$, $g_2$, $d_1$, $d_2$, $s_b$, and $s_{\rm{cut}}$ are free parameters, and $x$ and $f(x)$ denote the bin center and content of the image size histogram. Figure~\ref{fig:size_dist_cluster} shows the distribution and the resulting fitted functions for several example cleanings.

\begin{figure*}[h!]
    \centering
    \includegraphics[width=\linewidth]{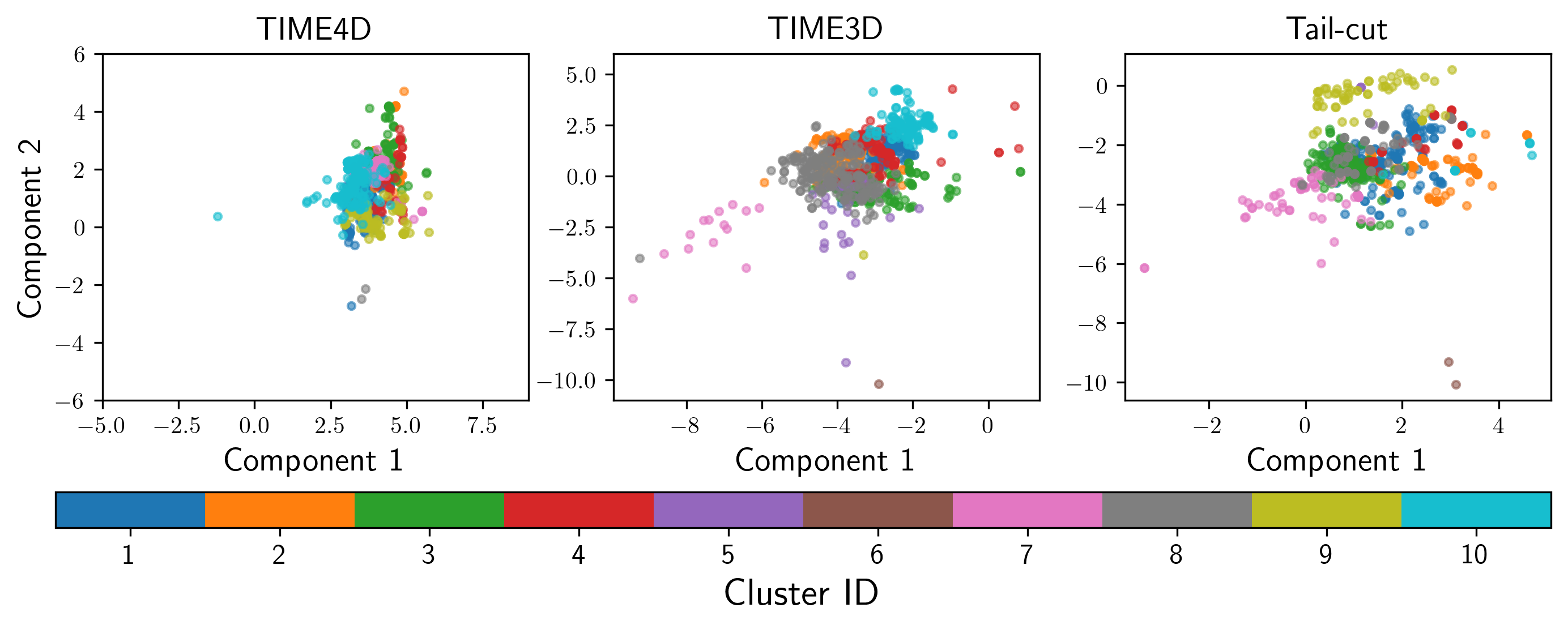}
    \caption{Results of K-means clustering for three different cleaning algorithms:\texttt{TIME3D}, \texttt{TIME4D}, and tail-cut. Ten cluster regions are estimated for each algorithm. This approach reduces the parameter space to ten per algorithm, with each cluster treated as a single point. For visualization, PCA  is performed and the 14-dimensional parameter space is reduced to two components (Component 1 and Component 2).}
    \label{fig:groupclusters}
\end{figure*}

The seven fit parameters for gamma MC and seven for background data capture key aspects of the cleaned distributions, such as the total number of events remaining after cleaning, break points, and size thresholds. Cleanings with similar fitted parameters are therefore expected to result in similar experimental performances (e.g., sensitivity). Thus, to classify similar candidates, a clustering method was adopted. A K-means clustering algorithm~\citep{IKOTUN2023178} was trained for each cleaning method, resulting in ten clusters for each cleaning, as shown in Fig.~\ref{fig:groupclusters}. 

For visualization purposes only, a principal component analysis (PCA) ~\citep{MACKIEWICZ1993303} was applied to obtain two representative linear combinations of the 14-D size parameter space. Figure \ref{fig:groupclusters} illustrates the ten cluster regions for each cleaning algorithm. 

The candidate closest to each cluster center was chosen as representative of that cluster. These ten candidates serve as typical exemplars of their groups and were selected for further sensitivity evaluation. In addition to the cluster representatives, a further 30 (ten from each tested cleaning algorithm) were randomly selected to allow a more detailed examination of the parameter space. A total of 64 candidates–including the 30 cluster representatives (e.g., \texttt{TIME3D\_1} to \texttt{TIME3D\_10}), the 30 random selections (e.g., \texttt{TIME3D\_11} to \texttt{TIME3D\_20}), and the four default cleanings (\texttt{0916NN2}, \texttt{0714NN2}, \texttt{TIME3D\_STD1}, \texttt{TIME3D\_STD2})– are included in the next step of this workflow. This clustering approach not only reduces the tested parameter space to a manageable number, but also preserves diversity in the cleaning configurations.

\subsection{Performance evaluation}

Gamma-ray energy and direction reconstruction as well as gamma-hadron separation were performed for all 64 candidates according to the procedure of \cite{unbehaun2025improvementsmonoscopicanalysisimaging}, which introduced several improvements to the low-energy end of CT5 mono performance.

The training procedure, implemented within the HAP framework, follows the steps outlined below (further details can be found in \cite{unbehaun2025improvementsmonoscopicanalysisimaging}). First,
preselection cuts are applied, requiring an image size above 50 p.e. and more than five pixels. Next, gamma-ray reconstruction training is performed using neural networks that are trained to flip images and reconstruct both direction and energy. A gamma-hadron separation is achieved by training a BDT with gamma MC and off-data events. A size-dependent BDT cut is then optimized to maximize the q-factor, defined as $q=\epsilon_{\mathrm{sig}}/ \sqrt{\epsilon_{\mathrm{bkg}}}$ , and is subsequently smoothed.

\subsubsection{Improvement in sensitivity and optimization of preselection cuts}

Training performance at larger sizes is independent of whether smaller images are included, which justifies the use of very loose preselection cuts prior to training. However, the final performance is strongly dependent on these cuts. Loose cuts allow for a lower energy threshold as more faint images are retained. However, high-energy events with large impact parameters also result in faint images, which are harder to reconstruct. As a result, looser preselection cuts deteriorate performance across the whole energy range. Consequently, the final preselection cuts are optimized by estimating the improvement in sensitivity with respect to the loosest case, i.e., 50 p.e. and five pixels. Similarly to~\cite{HASSAN201776}, we evaluate this improvement, denoted $\xi$, as 

\begin{equation} \label{eq:sens_improvement}
    \xi = \left ( \frac{ \sum_i^N \frac{S_i^{\rm{(50pe-5pix)}}}{S_i^{\rm{(cut)}}}}{N} \right)^{-1},
\end{equation}

where $S_i$ denotes the differential sensitivity in each pixel. All sensitivity calculations are performed assuming point-like gamma-ray sources, in line with standard IACT analysis procedures. For each candidate, we consider two criteria.

The performance criterion is evaluated by calculating $\xi$ for bins between $50\,$GeV and $500\,$GeV. This evaluates the broad-energy performance of the cleaning in both spectral and spatial analysis in which good sensitivity is required throughout the whole energy range. The upper limit of $500\,$GeV is chosen because stereoscopic analyses are expected to perform better in this range.
 The detection criterion is evaluated by calculating $\xi$  for bins between $36.5\,$GeV and $86.5\,$GeV. This evaluates the capability of achieving the detection of faint fluxes at low energies. Extending the analysis to lower energies results in reduced spectral resolution over the entire energy range. Therefore, this criterion focuses primarily on being able to detect faint sources, such as transients, at the expense of reconstruction accuracy for these events.

\subsubsection{Fake cluster rate}
As a consistency check, we estimated the fake cluster rate for all 64 candidates. We used pure NSB simulations, varying the NSB scaling factor $x_{\rm{nsb}}$, such that $x_{\rm{nsb}} \times $the realistic NSB rate was cleaned and the number of images kept after cleaning was estimated. The realistic NSB rate is defined as the mean of the Gaussian distribution fitted to all observations, excluding extreme NSB conditions. 
Candidates with a fake cluster rate above $1\%$ in any of the realistic NSB simulation scenarios considered ($x_{\rm{nsb}}=1.0$ and $x_{\rm{nsb}}=1.645$) were discarded. Twenty-four candidates were affected by this limit, which resulted in $36.9\%$ of all tested candidates being rejected. This selection ensured that the cleaning algorithms did not retain images without any gamma-ray-induced signal and, in particular, do not retain secondary noise clusters in the gamma images.

\subsubsection{Effective area ratio between nominal and high NSB simulations}

\begin{figure}[h]
    \centering
    \includegraphics[width=\linewidth]{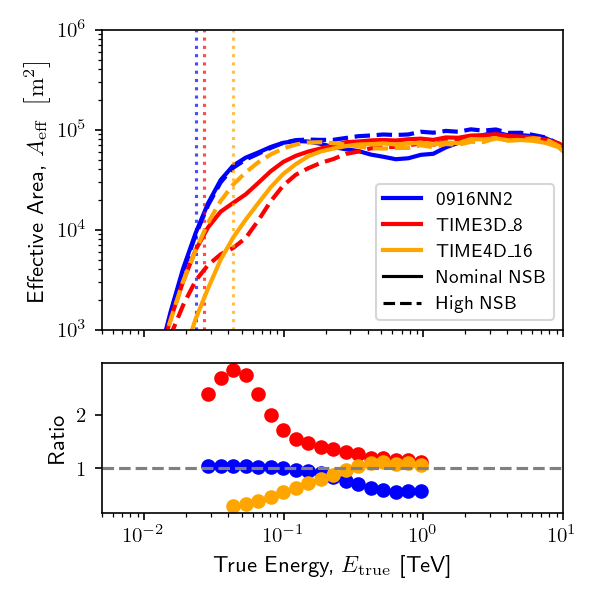}
    \caption{Effective area for representative cleaning candidates calculated from nominal (full lines) and high NSB simulations (dashed lines). Horizontal dashed lines indicate the energy threshold, defined as the energy at which the effective area is $10\%$ of its maximum. The bottom panel shows the ratio between the nominal and high NSB cases. Both example time-cleaning candidates illustrate extreme cases of overestimating or underestimating the signal efficiency at higher NSB rates; both are discarded according to the set limit.}
    \label{fig:Aeff ratio}
\end{figure}

To quantify the impact of NSB on performance, we computed the effective areas, including the optimized preselection cuts and gamma-hadron separation,  as a function of the MC energy for both nominal and high NSB simulations for all 64 candidates. Figure \ref{fig:Aeff ratio} presents the effective areas for a representative candidate from each cleaning method, each optimized using its respective detection cut configuration. For NSB-susceptible cleaning candidates, the differences between the two NSB levels can vary in both directions, resulting in either the retention of a fake signal (orange) or misclassification of the signal as background (red). We calculated the average ratio ($\kappa$) between the effective areas for the nominal and high NSB cases for energy bins between the energy threshold and 1 TeV. The threshold is defined as the energy for which the effective area is $10\%$ of its maximum and 1 TeV is chosen as it represents the primary energy regime for CT5-only operations. To keep NSB systematics for changing NSB levels below 10$\%$, the ratio should not exceed $1.1$.

\section{Results}

\begin{figure*}[h!]
    \centering
    \includegraphics[width=\linewidth]{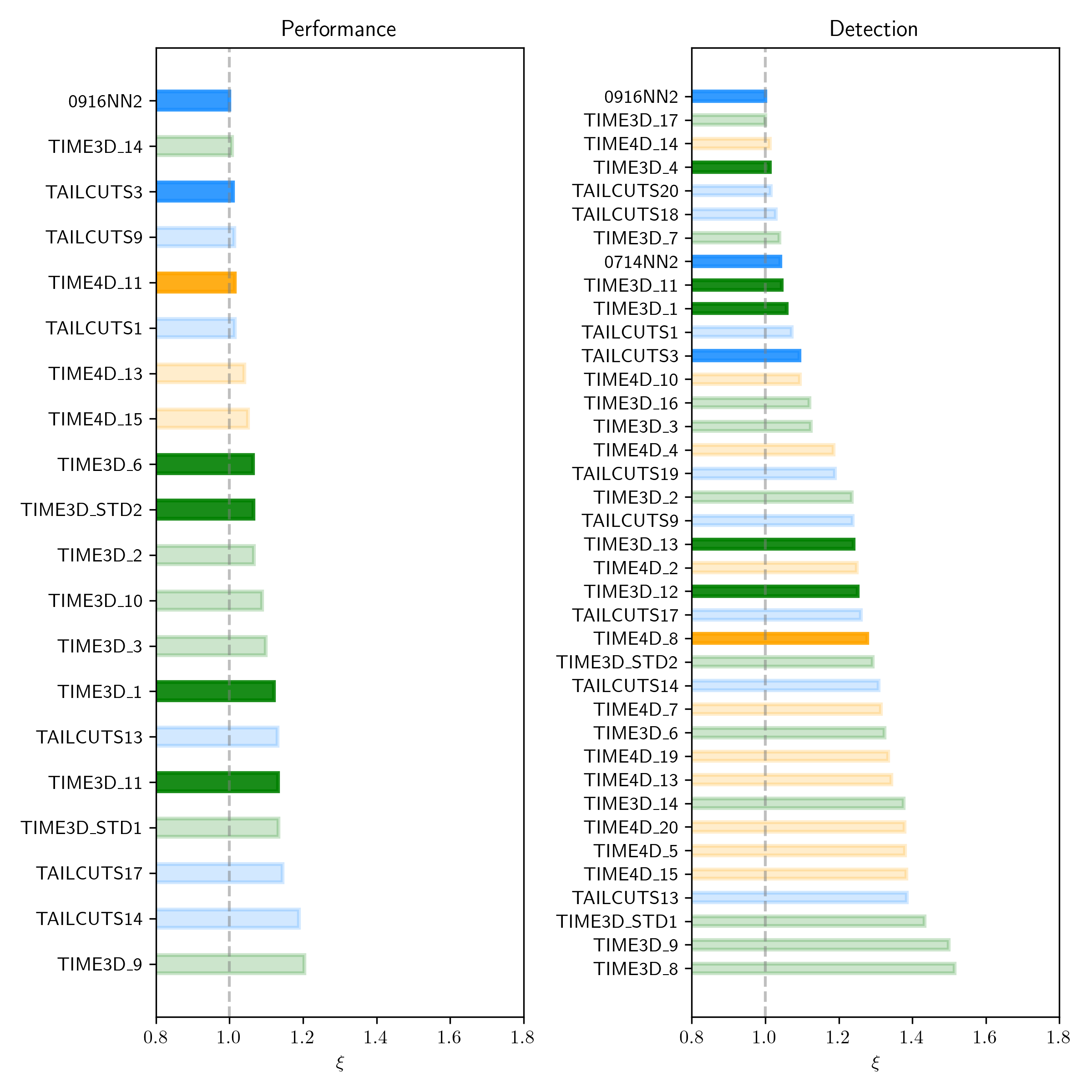}
    \caption{Sensitivity improvement for candidates with better performance that the default tail-cut  cleaning (\texttt{0916NN2}) in the performance range ($50\,$GeV to$500\,$GeV) and the detection range ($36.5\,$GeV to $86.5\,$GeV). Candidates that fail the fake cluster rate and/or the average effective area ratio criteria are shown as lighter colors. Blue, green, and orange bars represent candidates for tail-cut, \texttt{TIME3D}, and \texttt{TIME4D,} respectively.}
    \label{fig:PPUT}
\end{figure*}

\begin{figure*}[h!]
    \centering
    \includegraphics[width=0.49\linewidth]{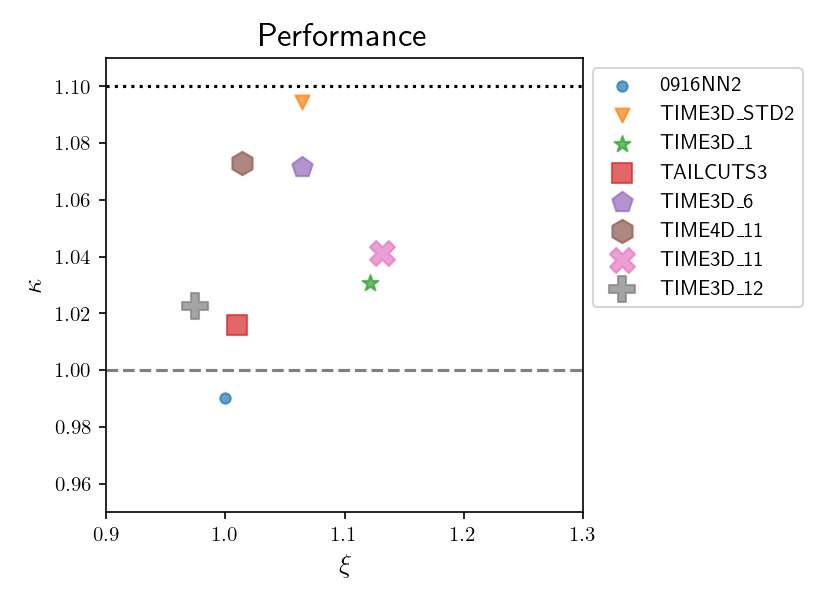}
    \includegraphics[width=0.49\linewidth]{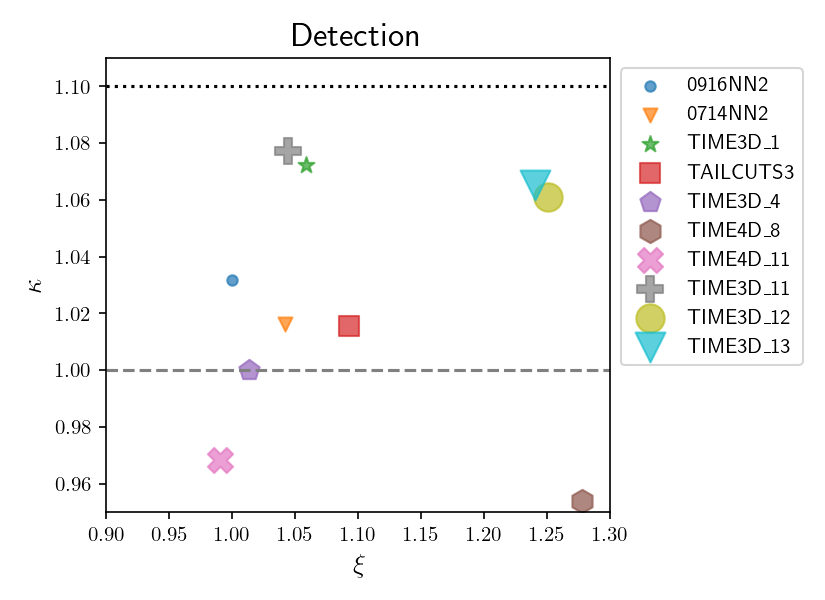}
    
    \caption{Average improvement in sensitivity, $\xi$, versus average effective area ratio, $\kappa$, for the best candidates. The optimal region is as close as possible to $\kappa$ of 1 and at large values of $\xi$.}
    \label{fig:pput_aeff}
\end{figure*}

The application of the optimization pipeline yields the optimized sensitivity for each of the 64 candidates from the three image-cleaning algorithms for IACTs: \texttt{TIME3D}, \texttt{TIME4D}, and the conventional tail-cut method.
Figure \ref{fig:PPUT} compares the representative candidates based on their improvement in sensitivity, $\xi$, under both the performance and detection criteria. Only those representatives that demonstrate better performance than the reference cleaning (\texttt{0916NN2}) are shown. Many candidates with large improvements are discarded due to the fake cluster rate and/or the average effective area ratio tests, i.e., they are too susceptible to NSB. These candidates are shown in transparent colors. As expected, it is easier to achieve apparently better performance when NSB susceptibility is not property controlled. Figure~\ref{fig:pput_aeff} shows the average effective area ratio for the remaining candidates, i.e., those with $\xi > 1$ that pass the NSB susceptibility requirements. An optimal candidate presents an average effective area ratio $\kappa$ of 1 and as large a $\xi$ as possible. Consequently, the best candidates are \texttt{TIME3D\_1} for performance and \texttt{TIME4D\_8} for detection. Their parameters and preselection cuts, as well as the best preselection cuts for \texttt{0916NN2} in each criterion, are listed in Table~\ref{tab:best}. A comparison of the Hillas parameters is provided in Appendix \ref{Appendix:hillas}.

\begin{table*}
    \caption{Preselection cuts and input parameters for the best candidate, and best preselection cuts for standard tail-cuts cleaning for each criterion.}
    \label{tab:best}
    \centering
    \begin{tabular}{c|c|c|c|c|c|c|c|c|c}
    \hline
    \hline
    \multirow{2}{*}{Criterion} & \multirow{2}{*}{Candidate} & \multicolumn{2}{c|}{Preselection cuts} & \multicolumn{6}{c}{Input parameters} \\
    & & Size [p.e.] & \# pixels & $n_{\rm{hard}} \,$[p.e.] & $n_{\rm{noise}}\,[\sigma_{\rm{noise}}]$ & minPts & $s_{\rm{scale}} \,$[m] & $t_{\rm{scale}} \,$[ns] & $A_{\rm{scale}}$ \\
    \hline
    \hline
    \multirow{2}{*}{Performance} & \texttt{0916NN2} & 275 & 5 & - & - & - & - & - & - \\
     & \texttt{TIME3D\_1} & 275 & 5 & 3 & 1 & 9 & 0.3 & 0.75 & - \\
     \hline
    \multirow{2}{*}{Detection} & \texttt{0916NN2} & 50 & 5 & - & - & - & - & - & - \\
     & \texttt{TIME4D\_8} & 95 & 7 & 3 & 7.2 & 3 & 0.1 & 4.16 & -2.67 \\
     \hline
     \hline
    \end{tabular}
\end{table*}

\subsection{Impact on Instrument Response Functions}

A closer examination of the instrument response functions (IRFs) provides important insights into the key differences that lead to improved performance among the new candidates. We evaluate performance after gamma–hadron separation in terms of four key metrics: effective area, angular resolution, energy bias, and energy resolution. While these quantities are derived from the IRF, they are not IRFs themselves. The IRFs describe the transformation between true and reconstructed gamma-ray properties and are factorized into components such as the effective collection area, the point spread function (PSF), and energy dispersion. From these, we extract scalar performance metrics –for example, angular resolution defined as the 68\% containment of the PSF, and energy resolution as the 68\% width of the reconstructed-to-true energy ratio distribution.

\subsubsection{Best candidate for the performance criterion}

\begin{figure*}[h!]
    \centering
    \includegraphics[width=\linewidth]{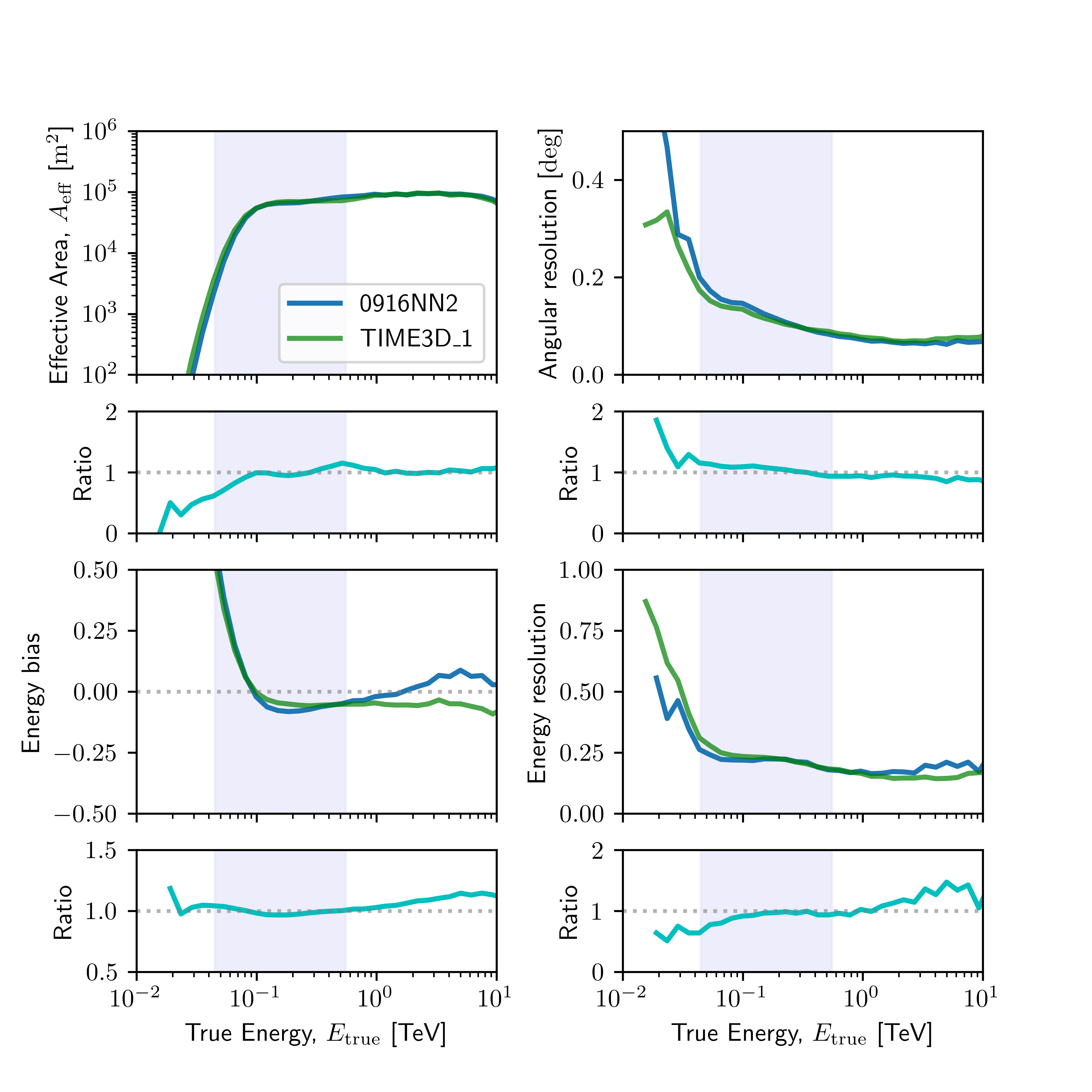}
    \caption{Comparison of the instrument response functions (effective area, angular resolution, energy bias, and energy resolution) between the current default tail-cut cleaning and the best candidate for the performance criterion, \texttt{TIME3D\_1}.The blue shaded region indicates the energy range of interest for the performance criterion.}
    \label{fig:irfs_performance}
\end{figure*}

\begin{figure}[h]
        \centering
        \includegraphics[width=\linewidth]{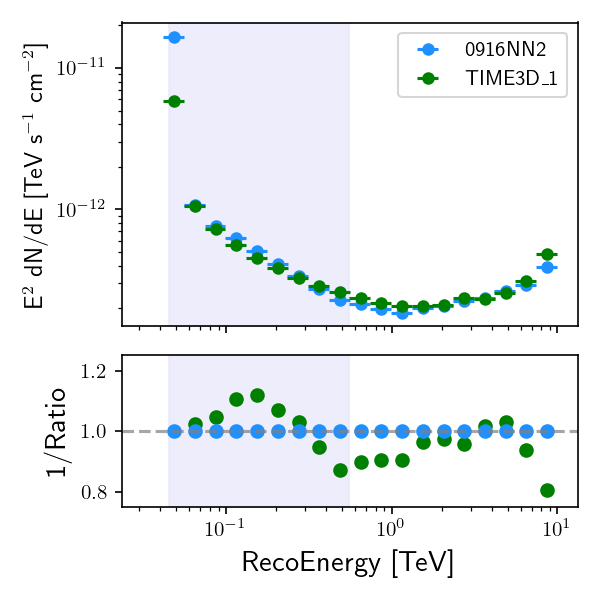}
        \caption{Differential sensitivity comparison between the current default tail-cut cleaning candidate and the best candidate for the performance criterion, \texttt{TIME3D\_1}. The upper panel shows the curves, while the lower panel displays the inverse ratio (i.e., values larger than 1 denote a smaller, and thus improved, sensitivity). The first flux point for \texttt{TIME3D\_1} at $E= 48.6\,\mathrm{TeV}$ corresponds to an improvement of $200\%$ and is omitted from the lower panel for clarity. The blue shaded region indicates the energy range of interest for the performance criterion.}
        \label{fig:Sensitivity-performance}
\end{figure}

Figure \ref{fig:irfs_performance} shows a comparison of the IRFs of \texttt{TIME3D\_1} and \texttt{0916NN2}. The effective area slightly increases at lower energy ranges, suggesting improved detection capability for low-energy gamma rays. Additionally, \texttt{TIME3D\_1} improves direction reconstruction, as indicated by the angular resolution. This improvement is important for maintaining more signal events. Energy reconstruction remains similar, with a reduced bias at the lowest energies but slightly worse energy resolution. 
Figure~\ref{fig:Sensitivity-performance} provides a detailed view of the sensitivity improvements for \texttt{TIME3D\_1}. It reveals a significant enhancement, particularly at the first flux point, where a factor of two improvement is observed, and for energies up to 300 GeV, where a 10-15\% improvement is expected. These gains are primarily driven by its improved angular resolution and greater gamma-ray retention. In the energy range around 1 TeV, \texttt{TIME3D\_1} shows slightly worse sensitivity than \texttt{0916NN2}. However, in this energy range, the smaller telescopes provide better sensitivity, and CT5 in monoscopic mode becomes less critical.

\subsubsection{Best candidate for the detection criterion}

\begin{figure*}[h!]
    \centering
    \includegraphics[width=\linewidth]{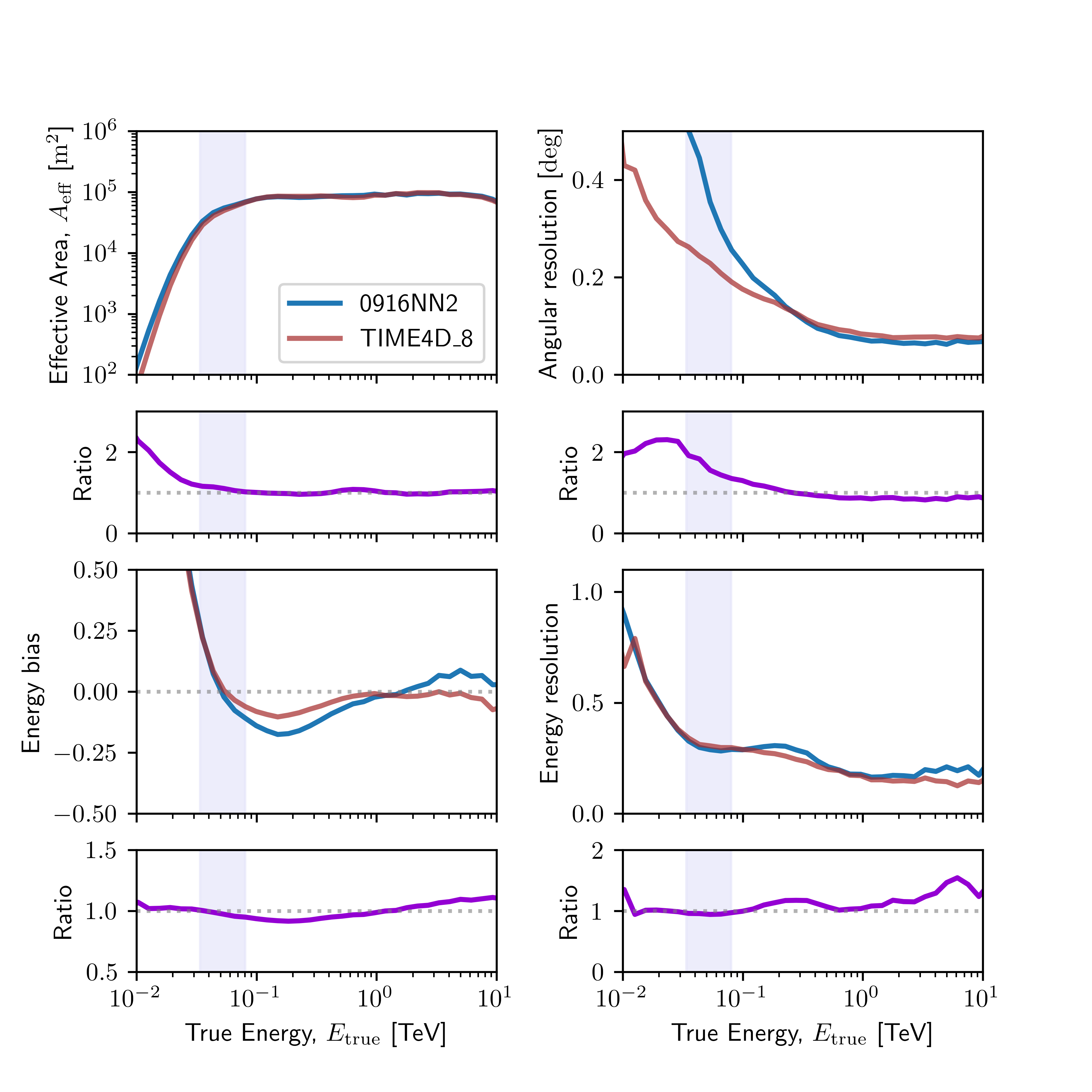}
    \caption{Comparison of the instrument response functions (effective area, angular resolution, energy bias, and energy resolution) between the current default tail-cut cleaning and the best candidate for the detection criterion, \texttt{TIME4D\_8}. The blue shaded region indicates the energy range of interest for the detection criterion.}
    \label{fig:irfs_detection}
\end{figure*}

\begin{figure}[h!]
        \centering
        \includegraphics[width=\linewidth]{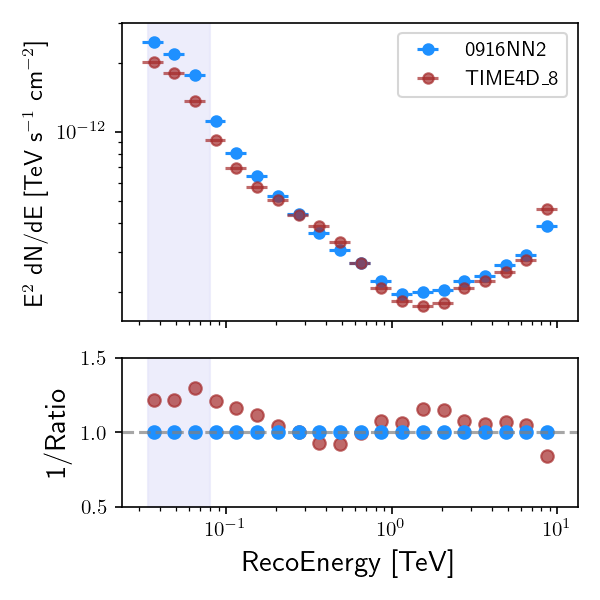}
    \caption{Comparison of the differential sensitivity for the current default tail-cut cleaning candidate and the best candidate for the detection criterion, \texttt{TIME4D\_8}. The upper panel shows the curves, while the lower panel shows the inverse of the ratio (i.e., values greater than 1 denote a smaller, and thus improved, sensitivity). The blue shaded region indicates the energy range of interest for the detection criterion.}
    \label{fig:Sensitivity}
\end{figure}

Time-based cleaning methods such as \texttt{TIME3D\_1} outperform the default cleaning approach in terms of event reconstruction accuracy,  supporting their effectiveness in mitigating noise while preserving critical shower information.
However, when the primary goal is detection optimization, different considerations apply. 
Figure~\ref{fig:irfs_detection} shows the IRFs for the standard cleaning \texttt{0916NN2} with preselection cuts optimized for detection (50 p.e., five pixels) and the best detection candidate, \texttt{TIME4D\_8}, using preselection cuts of 95 p.e. and seven pixels.
As discussed previously, an improved detection range can be achieved for a given cleaning method simply by loosening its preselection cuts. This can be seen in the effective area, which is slightly larger for \texttt{0916NN2} at the lowest energies. However, such loosening leads to poorer reconstruction performance. An optimal cleaning for the detection criterion should ensure that the reconstruction is not significantly degraded by loosening the preselection cuts. This becomes clear in the angular and energy reconstructions. \texttt{TIME4D\_8} maintains a much better angular resolution and energy bias compared to \texttt{0916NN2}, while still reaching a similar energy threshold. The resulting sensitivities are shown in Fig.~\ref{fig:Sensitivity}. The time-cleaning candidate yields an approximate $20\% $ improvement in sensitivity at low energies compared to the standard tail-cut cleaning.

The choice between the two best candidates, \texttt{TIME3D\_1} and \texttt{TIME4D\_8}, depends on the specific scientific objectives. If the goal is to maximize overall sensitivity across all energies, such as in spectral and spatial studies, \texttt{TIME3D\_1} provides the best solution. However, for optimizing low-energy event detection, such as the search for transients or faint sources, \texttt{TIME4D\_8} is more advantageous due to its superior noise suppression and enhanced event reconstruction at low energies.

\section{Conclusion}
This study introduces both a new time-based cleaning technique for H.E.S.S. and a comprehensive optimization pipeline to evaluate the image cleaning performance of IACTs. The method presented here accounts for correlations in gamma-ray induced air showers in both the pixel amplitude and timing information. We explored two options: \texttt{TIME3D}, which clusters pixel position and time; and \texttt{TIME4D}, which adds an extra dimension based on pixel amplitude relative to the brightest pixel. By systematically evaluating and refining the cleaning configurations of three algorithms –the tail-cut method, \texttt{TIME3D}, and \texttt{TIME4D} –this work addresses the challenge of balancing signal retention and noise suppression to improve overall sensitivity. Unlike traditional optimization approaches that focus on isolated metrics such as size retention or noise reduction, our method integrates multiple factors to ensure holistic performance improvement, focusing on sensitivity gain. Key innovations include an NSB susceptibility assessment using a BDT model to identify stable cleaning configurations under varying NSB conditions; an efficient parameter space exploration using K-means clustering to systematically group and evaluate diverse cleaning parameter sets while reducing computational costs; and a sensitivity-driven optimization approach that prioritizes the impact on key instrument response functions, including gamma-hadron separation, energy reconstruction, and effective area.

We demonstrate that time-based cleaning techniques significantly outperform the conventional tail-cut method. We explored two possible performance methods.  For overall improvement over the entire energy range, the best candidate was identified using the \texttt{TIME3D} method. It provides stable performance across a broad energy range, effectively balancing NSB suppression and signal retention, making it particularly useful for general-purpose gamma-ray analyses where uniform sensitivity is required. To improve the detection of faint sources at the lowest energies, \texttt{TIME4D} provides the best candidate. It achieves an approximately $ 25\%$ improvement in sensitivity at low energies by more aggressively filtering NSB noise while preserving enough shower structure for precise reconstruction. This enables looser preselection cuts while maintaining reasonable reconstruction resolution.

A key insight from this work is that simply expanding the effective area at low energies—for example by relaxing preselection cuts—does not necessarily lead to improved sensitivity. While this may increase the event rate, it can also severely degrade the reconstruction quality, particularly for direction and energy. Because sensitivity is computed under the assumption of point-like gamma-ray sources, accurate angular reconstruction is critical for separating signal from background. In this context, precise event reconstruction, especially in terms of angular resolution and gamma-hadron separation, becomes a dominant factor in achieving performance gains. For instance, the improved noise suppression of \texttt{TIME4D} enables significantly enhanced sensitivity, despite slightly higher energy thresholds. This highlights the importance of considering all aspects of performance, beyond simple metrics such as image size retention, when optimizing image-cleaning strategies.

Although validated here for H.E.S.S., this methodology is directly applicable to both current and future IACTs, such as VERITAS, MAGIC, and CTAO. Similar approaches can also be extended to other high-energy astrophysics experiments. By refining image-cleaning strategies, this work enhances our ability to detect faint astrophysical sources and, consequently, to better understand the most energetic phenomena in the Universe. Furthermore, the results highlight the necessity of tailoring image-cleaning algorithms to specific observational goals.

\begin{acknowledgements}
We thank the H.E.S.S. Collaboration for providing the simulated data, common analysis tools, and valuable comments on this work. 
\end{acknowledgements}

\bibliographystyle{aa}
\bibliography{references}

\newpage
\begin{appendix}
\section{Distribution of Hillas parameters} \label{Appendix:hillas}
We show the distribution of Hillas variables. Depending on the used cleaning candidate and preselect cut configuration, $\approx$ 130000 - 70000 background events, taken from observations where known gamma-ray sources are excluded.  All simulations were done for 20 deg zenith angle, 0 deg azimuth and 0.5 deg offset angle and the simulated spectrum with an index of -2 is re-weighted to -2.5. The distributions are normalized to 1 such that the difference in event numbers is not visible which means the y-axis of the plots shows the normalized number of entries in that bin. The upper panel always compares the default cleaning with \texttt{TIME3D\_1} after applied optimized pre-selection performance criteria cuts from \ref{tab:best}, the lower one always compares the default with \texttt{TIME4D\_8} after applied detection criteria cuts. The solid line shows the distribution of the MC gamma simulation, the dashed one for the real background data.

\begin{figure}[h!]
    \centering
    \includegraphics[width=0.9\linewidth]{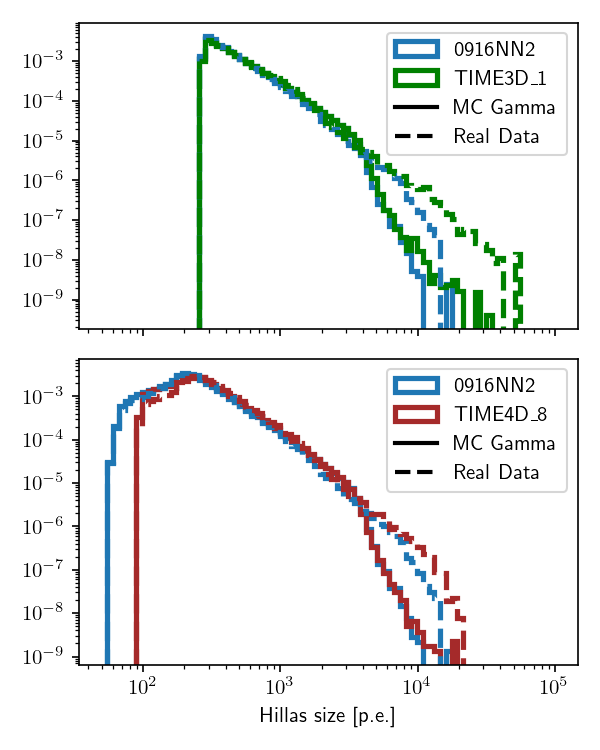}
     \caption{size distribution of \texttt{0916NN2}, \texttt{TIME3D\_1} and \texttt{TIME4D\_8}.}
    \label{fig:enter-label}
\end{figure}

\begin{figure}[h!]
    \centering
    \includegraphics[width=0.9\linewidth]{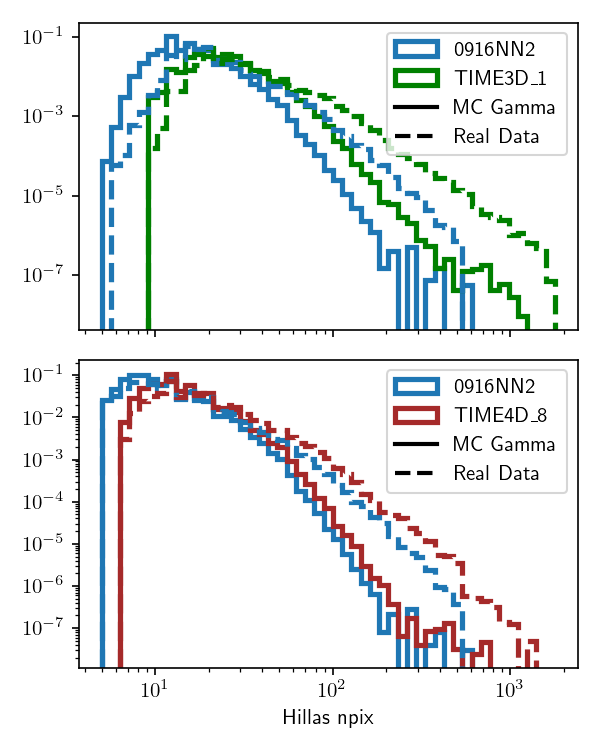}
    \caption{Number of remaining pixels distribution of \texttt{0916NN2}, \texttt{TIME3D\_1} and \texttt{TIME4D\_8}.}
    \label{fig:enter-label}
\end{figure}

\begin{figure}[h!]
    \centering
    \includegraphics[width=0.9\linewidth]{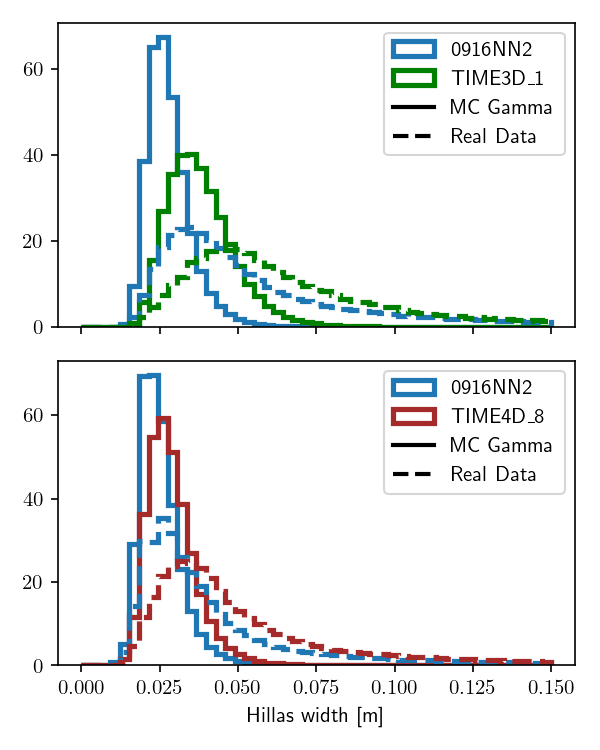}
    \caption{Hillas width distribution of \texttt{0916NN2}, \texttt{TIME3D\_1} and \texttt{TIME4D\_8}.}
    \label{fig:enter-label}
\end{figure}

\begin{figure}[h!]
    \centering
    \includegraphics[width=0.9\linewidth]{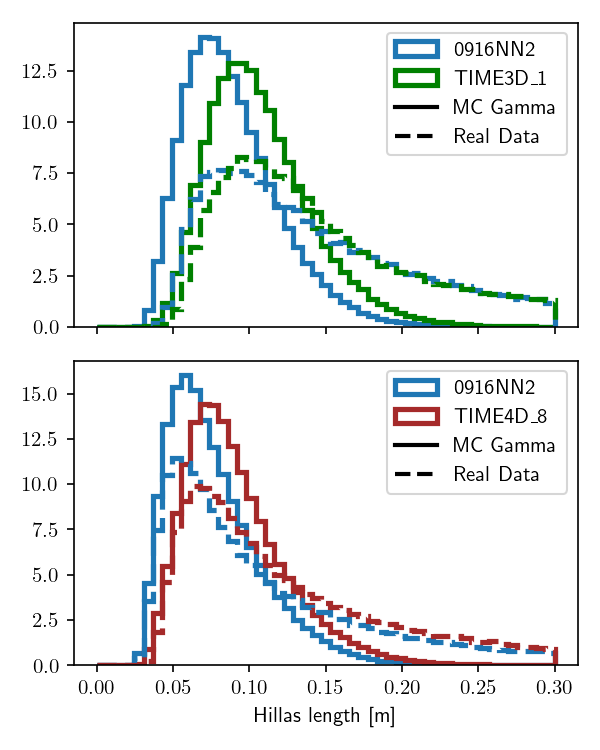}
    \caption{Hillas length distribution of \texttt{0916NN2}, \texttt{TIME3D\_1} and \texttt{TIME4D\_8}.}
    \label{fig:enter-label}
\end{figure}

\begin{figure}
    \centering
    \includegraphics[width=0.9\linewidth]{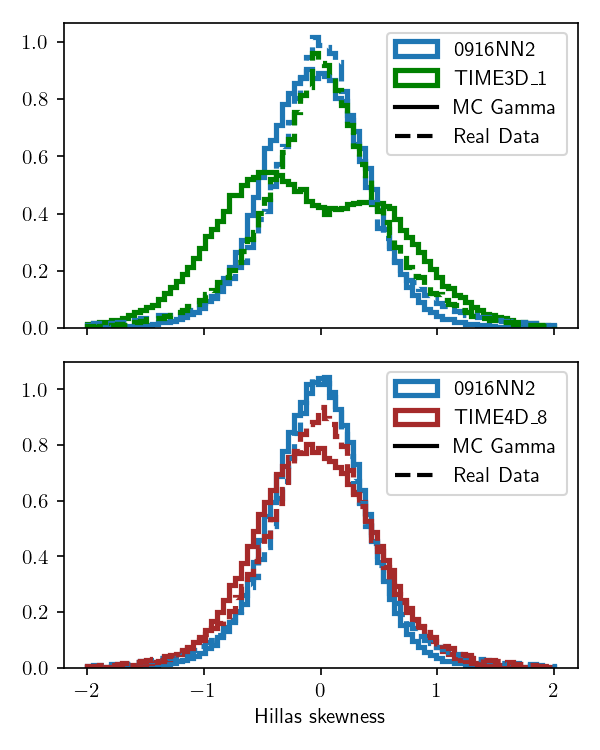}
    \caption{Hillas skewness distribution of \texttt{0916NN2}, \texttt{TIME3D\_1} and \texttt{TIME4D\_8}.}
    \label{fig:enter-label}
\end{figure}

\begin{figure}[h!]
    \centering
    \includegraphics[width=0.9\linewidth]{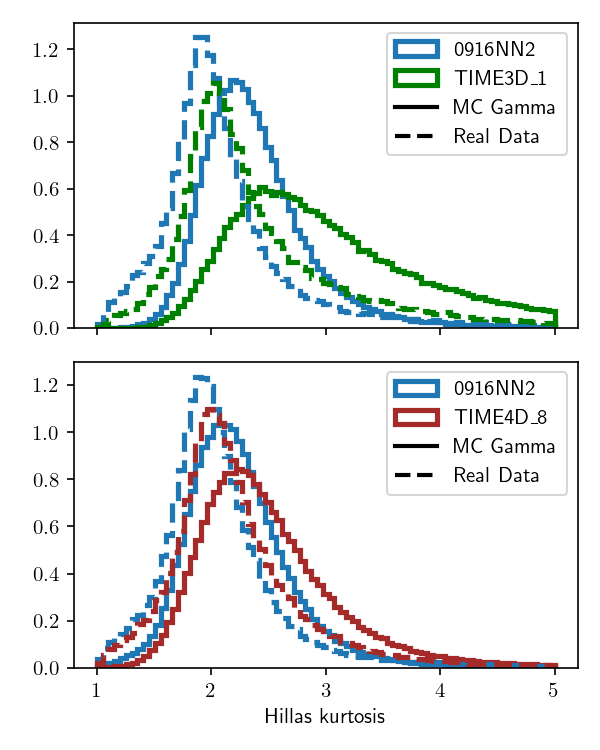}
    \caption{Hillas kurtosis distribution of \texttt{0916NN2}, \texttt{TIME3D\_1} and \texttt{TIME4D\_8}.}
    \label{fig:enter-label}
\end{figure}

\begin{figure}[h!]
    \centering
    \includegraphics[width=0.9\linewidth]{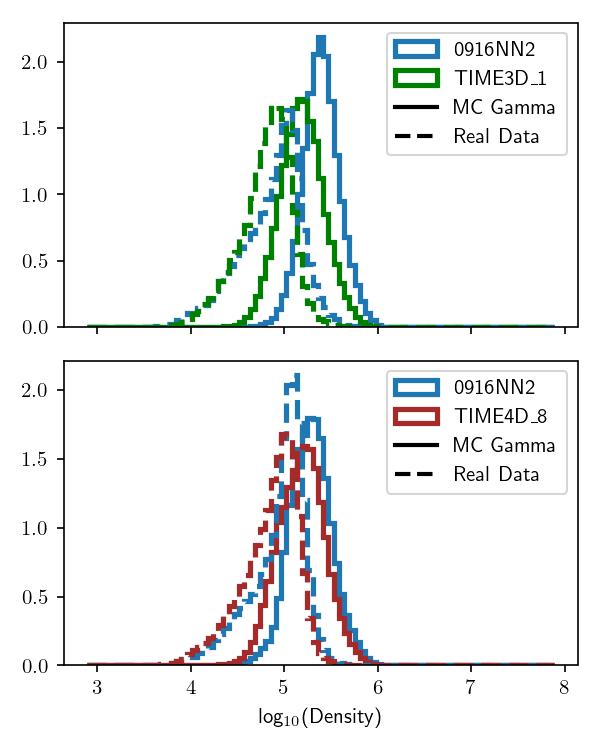}
    \caption{Density distribution of \texttt{0916NN2}, \texttt{TIME3D\_1} and \texttt{TIME4D\_8}.}
    \label{fig:enter-label}
\end{figure}

\begin{figure}[h!]
    \centering
    \includegraphics[width=0.9\linewidth]{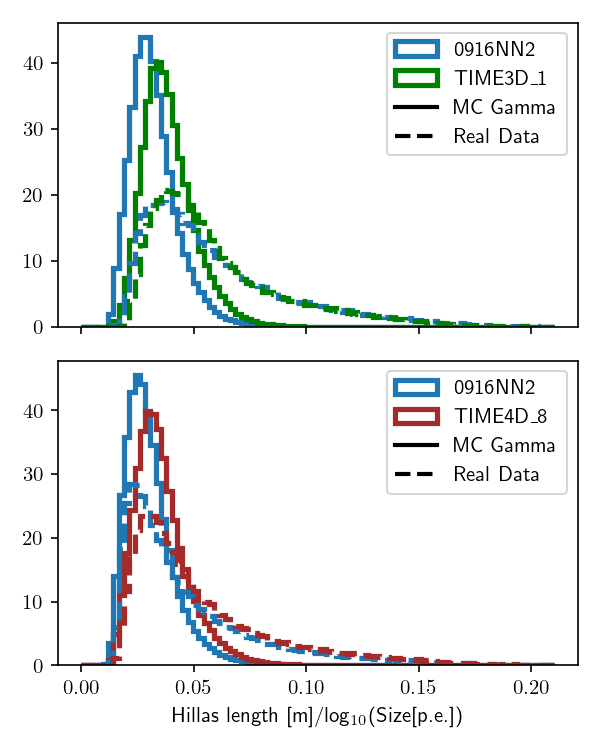}
    \caption{Hillas length over size distribution of \texttt{0916NN2}, \texttt{TIME3D\_1} and \texttt{TIME4D\_8}.}
    \label{fig:enter-label}
\end{figure}
\end{appendix}
\end{document}